# A Simplified Model for the Battery Ageing Potential Under Highly Rippled Load for Battery Management and Active Degradation Control

T. Kacetl, J. Kacetl, N. Tashakor, and S. M. Goetz

*Abstract*—Whereas in typical standardized tests batteries are almost exclusively loaded with constant current or relatively slowly changing cycles, real applications involve rapid load ripple, which do not contribute to the net energy. The trend to reduced filter capacitors and even dynamically reconfigurable batteries further increases the ripple. The influence of rippled load on lithium batteries is therefore receiving increased attention. According to recent studies, accelerated ageing strongly depends on the frequency of the ripple. We use electrochemical models to derive a highly simplified regression model that catches the asymptotic behavior and allows parameter identification and calibration to specific cells. The model allows quantitative monitoring of the additional ageing due to ripple current in battery management systems. Furthermore it enables active control of the ageing potential by influencing the frequency content in modern battery systems, such as reconfigurable batteries.

## I. INTRODUCTION

High power and energy density stimulates the use of lithium-ion batteries in a wide range of applications, such as portable devices, battery energy storage systems (BESS), and electric vehicles [1]. During their life time, materials of the cell components undergo constant degradation associated with chemical and structural changes, which results either in loss of lithium inventory, loss of active material, or increase of the cell impedance [2]. In addition to ambient temperature and state of charge (SoC), which plays an exclusive role in calendar ageing, character of the load shows to have dominant influence during cycling [3].

Demands on the battery load progressively increase with advances in semiconductor technology and scaling of power converters. Especially the automotive industry pushes for peak performance and high-power charging, where batteries experience loads of multiples of the C-rate (ampere–hour capacity) [4, 5]. The discharge rate and related increase of the cell temperature during high current load follow Arrhenius law [6, 7] per

$$y_k = A \exp\left(-\frac{E_a}{RT} + \alpha(k)(I_k - I_{1C})\right) k^z. \tag{1}$$

In addition, the conversion of energy in a traction inverter further introduces a wide range of harmonics into the battery load [8]. Although the battery side of the inverter is typically stabilized by a DC-link capacitor, cost and space constraints drive a trend to reduce them so that a large share of the ripple propagates to the battery [9]. Furthermore, alternative topologies such as battery-integrated modular multilevel converters and dynamically reconfigurable batteries generate even higher load ripple [10, 11]. Nevertheless, the ripple is commonly not considered in ageing models, and even cell manufacturers test and specify maximum current ratings for continuous load down to periods in second range, which is ripple-wise far from actual load conditions in automotive applications. Similarly, also battery models tend to approximate battery behavior for a direct current (DC) load and neglect dynamics of the battery in millisecond range [12].

Nevertheless, studies experimenting with rippled load are recently appearing [13]. Despite yet some inconsistent conclusions of individual studies regarding the influence of the load ripple on battery ageing, the frequency of the ripple proves to have unambiguous impact. The studies almost uniformly report vanishing additional ageing at higher frequencies, which is unanimously assigned to the double–layer and electrode capacitance [14]. Shunting behavior of the battery cells at higher frequencies leads to lower activation of faradaic processes at the electrode–electrolyte interface, and the high frequency ripple is supplied from charge accumulated at the interface[15]. Despite molecule-level modeling approaches based on a pseudo 2D model of the cell as presented in [16], there is not yet any Arrhenius-like model which would approximate the effect in a simple form and might even run online on a vehicle's battery management system.

We combine features of the 2D model with equivalent circuit modeling, which yields and verifies the simplified model approximating the ageing behavior and which allows identification and calibration to specific cell. The simplified model can be used to benchmark modulation techniques, selection of switching frequency and necessity of filter installation.

In Section II, we review battery modeling approach and electro-chemical processes in the battery cell. Section III presents a combined modeling approach, which forms a simple relation between frequency and ageing. Section IV as the experimental part of the manuscript reveals shunting behavior of the battery cell. Section V concludes the paper.

## II. BATTERY CELL MODELING

### A. Electrochemical model

Charging and discharging of battery cells is necessarily connected to (electro-)chemical activity and reactions, so-called faradaic processes. Through

$$C_6 + Li^+ + e^- \leftrightarrow LiC_6, \quad (2)$$

the lithium molecules are de-intercalated from one electrode, dissolved in the electrolyte, transported, and consequently intercalated into the counter electrode [17].

Intercalation on the electrode–electrolyte interface is governed by the Butler-Volmer (BV) equation per

$$j_{\text{int}}(\eta) = A\, j_{0,\text{int}} \left( \exp\left(\frac{\alpha_{\text{int}} F}{RT}(\eta - U_{\text{eq}}^{\text{int}})\right) - \exp\left(-\frac{(1-\alpha_{\text{int}})F}{RT}(\eta - U_{\text{eq}}^{\text{int}})\right) \right), \quad (3)$$

where
$j_{int}$ intercalation current in A;
$A$ surface of the porous electrode in m$^2$;
$j_{0,int}$ exchange current density in A m$^{-2}$;
$\alpha_{int}$ charge transfer coefficient;
$\eta$ over-potential in $V$;
$T$ temperature in K;
$F$ Faraday constant, $F = 96\,485$ C mol$^{-1}$;
$R$ gas constant, $R = 8.3145$ J mol$^{-1}$ K$^{-1}$;
$U_{\text{eq}}^{int}$ equilibrium potential in $V$;
and the over-potential $\eta$ represents difference of the electrochemical potential between solid electrode and liquid electrolyte phase. For small over-potentials ($\eta < 10$ mV), the equation can further be approximated and expressed as charge transfer resistance $R_{\text{CT}}$ per

$$j_{\text{int}}(\eta) = A\, j_{0,\text{int}} \frac{nF}{RT}\eta \quad (4)$$

$$R_{\text{CT}} \approx \frac{\eta}{j_{\text{int}}} = \frac{RT}{nFi_{0,int}}, \quad (5)$$

where n is the number of electrons participating in the reaction [18].

Except for the faradaic (intercalation) current, the electrode–electrolyte interface can also contribute to the load by accumulated charge at the interface. In contrast to faradaic processes, this source of charge immediately responds to any disturbance in load and progressively builds up local over-potentials necessary for the activation of the intercalation processes. The effect is modelled as a parallel plate capacitor $C_{\text{dl}}$, and this non-faradaic current can described as [12]

$$\begin{aligned} j_{\text{dl}} &= A\, C_{\text{dl}} \frac{\partial \eta}{\partial t}, \\ j_{\text{load}} &= j_{\text{int}} + j_{\text{dl}}. \end{aligned} \quad (6)$$

Further processes, such as ionic flux in the electrolyte are governed by Stefan–Maxwell diffusion in the electrolyte, Fick's second law diffusion in the solid electrode, etc. More details can be found in literature [17].

### B. Solid-eletrolyte interface layer and lithium plating

Aside from intercalation of the lithium in the electrode, especially the anode exhibits further chemical activity. Surface graphite electrodes react with the electrolyte and passivate the surface during initial cycles. The solid electrolyte interface layer (SEI) allows transport of lithium ions, but prevents further reaction with the electrolyte [19]. Nevertheless, growth of the SEI layer continues at a certain rate during lifetime of the battery cell. Equations (6) and (7) describe side reactions of electrolyte components—ethylene carbonate (EC) and dimethyl carbonate (DMC)—and production of key inorganic components of the SEI multilayer [20] per

$$\begin{aligned} 2Li^+ + 2(C_3H_4O_3)(EC) + 2e^- &\leftrightarrow (CH_2OCO_2Li)_2 + C_2H_4, \quad (7)\\ 2Li^+ + C_3H_6O_3(DMC) + 2e^- &\leftrightarrow Li_2CO_3 + C_2H_6. \quad (8) \end{aligned}$$

Both reactions happen at the same interface as the main reaction (3). Kinetics of the reactions and SEI growth is governed by cathodic BV per

$$j_{\text{SEI}}^{\text{EC}} = -F k_0^{\text{EC}} c_{\text{EC}}^s \left[\exp\left(-\frac{\alpha_{c,\text{EC}}F}{RT}(\eta - U_{\text{SEI}}^{\text{EC}})\right)\right], \quad (9)$$

$$j_{\text{SEI}}^{\text{DMC}} = -F k_0^{\text{DMC}} c_{\text{DMC}}^s \left[\exp\left(-\frac{\alpha_{c,\text{DMC}}F}{RT}(\eta - U_{\text{SEI}}^{\text{DMC}})\right)\right]. \quad (10)$$

At high over-potentials, the cell can additionally deposit lithium in metallic form directly onto surface of the graphite anode rather than intercalate into the lattice. Deposition of the metallic lithium, also called as lithium plating, is governed by

$$Li^+ + e^- \leftrightarrow Li_{(s)}, \quad (11)$$

$$j_{\text{pl}} = -i_0^{\text{pl}} \left[\exp\left(-\frac{\alpha_{c,\text{pl}}F}{RT}(\eta - U_{\text{eq}}^{\text{pl}})\right)\right]. \quad (12)$$

The side reactions are irreversible and assumed to be a major sources of ageing of the cell [21]. Lithium plating and SEI growth are typically irreversible reactions, which contribute to the loss of lithium inventory and cause capacity fade [22]. In addition, growth of the SEI layer hinders mass transport, increases electrical impedance of the battery cell, and lowers performance. Metallic lithium grows dendrites, which may in catastrophic scenarios puncture the separator and cause internal electric short circuits as well as thermal runaway of the battery cell [23].

### C. Randles' equivalent circuit

In contrast to complex electrochemical models, equivalent circuits of battery cells are widely used for their simplicity. Components of the cell, their properties, and internal chemical processes are approximated with combination of few electrical elements [24]. A reduced number of parameters further allows parameter identification with experimental data [25]. Figure 1 shows Randles' equivalent circuit, where both electrodes and intercalation processes are represented by a single charge-transfer resistance $R_{\text{CT}}$. Diffusion processes are generalized by the Warburg impedance $Z_{\text{W}}$, which can be further approximated by a string of R–C elements (see Figure 1) [26]. Charge in the capacitor $C_{\text{dl}}$ alternatively contributes to non-faradaic current and corresponds to electrode capacitance in Equation 6. Mass transportation limitations through the SEI layer are reflected in another R–C component, where resistance $R_{SEI}$ progressively increases as the cell ages. Resistance $R_0$ sums resistances of various components (electrolyte, contacts etc.), and inductance $L_s$ represents parasitic inductance of cells as well as electrical interconnection. Finally, the potential of electrodes is simulated by internal voltage source dependent on the state of charge (SoC).

While electrochemical models offer more insight in the chemical processes, parameter identification of the models is despite several attempts not feasible with conventional electrical measurements [27-29]. Nevertheless, description of the internal processes and their dependencies can be conveniently combined with easily identifiable equivalent circuit model to get a simple degradation model per our further suggestions.

### III. COMBINED MODELING APPROACH

Finding a simple form of frequency dependency of the cell ageing requires shrinking and generalization of the electrochemical model combined with properties of the equivalent circuit. Equations (9)–(12) describe loss of lithium inventory as a function of the over-potential, which simultaneously drives kinetics of the intercalation in the Butler–Volmer equation (3) and can be found in the equivalent circuit as charge transfer resistance $R_{CT}$ (see Figure 1).

Inspecting Equations 9–12, we assume an exponential relation between kinetics of the side reactions and over-potential. Since all corresponding reactions result in loss of lithium inventory, we sum them up as

$$j_{ageing} = j_{SEI}^{EC} + j_{SEI}^{DMC} + j_{pl} \qquad (13)$$

and further rewrite the individual equations in the form of

$$\begin{aligned}
j_{SEI}^{EC} &= -F k_0^{EC} c_{EC}^s \exp(U_{SEI}^{EC}) \left[\exp\left(-\frac{\alpha_{c,EC} F}{RT}\eta\right)\right], \\
j_{SEI}^{EC} &= -F k_0^{DMC} c_{DMC}^s \exp(U_{SEI}^{DMC}) \left[\exp\left(-\frac{\alpha_{c,DMC} F}{RT}\eta\right)\right], \\
j_{pl} &= -i_0^{pl} \exp\left(\frac{\alpha_{c,pl} F}{RT} U_{eq}^{pl}\right) \left[\exp\left(-\frac{\alpha_{c,pl} F}{RT}\eta\right)\right]. 
\end{aligned} \qquad (14)$$

The reaction rate of the considered side-reactions is an essential indicator of battery ageing processes, and we therefore introduce an ageing potential term, which benchmarks the reaction rates under specific load conditions. Per the suggestion in [21, 30], the charge coefficients of the side reactions are equal and we can substitute them with a common coefficient $\alpha_{c,ag}$ in Eqs. (13)–(14) as

$$\begin{aligned}
(\alpha_{c,EC} &= \alpha_{c,DMC} = \alpha_{c,pl}) = \alpha_{c,ag}, \\
j_{ageing} &= (k_{EC} + k_{DMC} + k_{pl}) \exp\left(-\frac{\alpha_{c,ag} F}{RT}\eta\right).
\end{aligned} \qquad (15)$$

As remarked earlier, we can find the electrochemical overpotential directly in the charge–transfer resistance of the equivalent circuit. Using Eqs. (5) and (15), the ageing rate can be written as a function of the intercalation current $j_{int}$ per

$$j_{ageing} = k_{ag} \exp\left(-\frac{\alpha_{c,ag} F}{RT} R_{ct} i_{int}\right). \qquad (16)$$

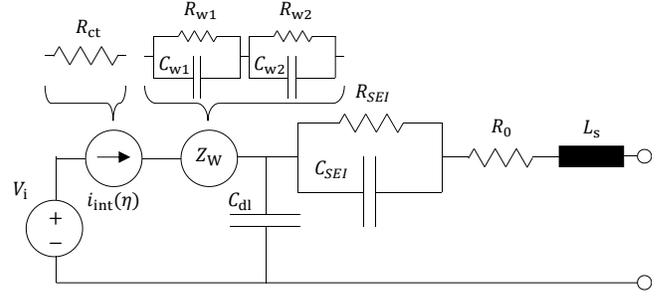

**Figure 1.** Randles' equivalent circuit modeling approach with charge transfer and diffusion approximation.

Considering DC load, the intercalation current $i_{int}$ is after all transients equal to the load current, which directly yields an exponential relation between ageing and load current, which can be cross validated to the Arrhenius law (1).

Nevertheless, during transients the electrode capacitance contributes to the load current according to Eq. (6) until the interfacial over-potential reaches high enough values to run intercalation reactions (ergo (dis-)charging current) at a sufficient rate. Similarly, the alternating current or any current ripple is partly supplied from the electrode capacitance. In principle, the battery cell offers two alternatives: a faradaic current related to chemical activity and a non-faradaic current related to charging of the electrode capacity.

The equivalent circuit (Figure 1) and impedance of the cell (Figure 2) represent this phenomenon under variable frequency. Electrode capacitance $C_{dl}$ acts as shunting path, which allows the flow of high–frequency components around the electrochemical reactions at minimum impedance and adds high–pass characteristic to the cell as Figure 2 outlines. The high impedance of diffusion processes limiting the electrochemistry at low frequencies is progressively bypassed by shunting of the load through the electrode capacitance, where the impedance reaches its minimum (given by electrolyte resistance and metallic contacting of electrodes). At higher frequencies, the impedance rises as the inductance of the cells dominates. Nevertheless, for a substantial part of the frequency range, the cell impedance can be approximated as

$$Z_{cell} \sim \frac{1}{\sqrt{f_c^2 + f^2}}, \qquad (17)$$

where $f_c$ corresponds to the cut-off frequency of the high pass characteristic. The magnitude of the AC component of the

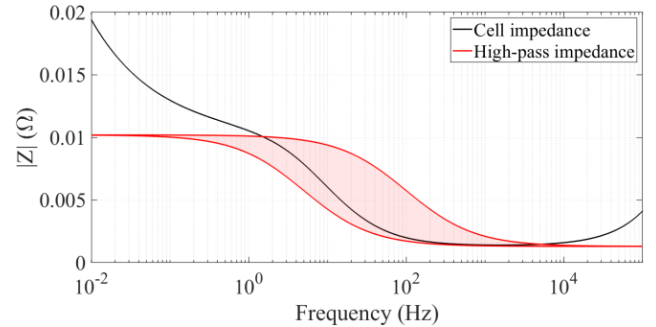

**Figure 2.** Cell impedance compared to high-pass characteristic.

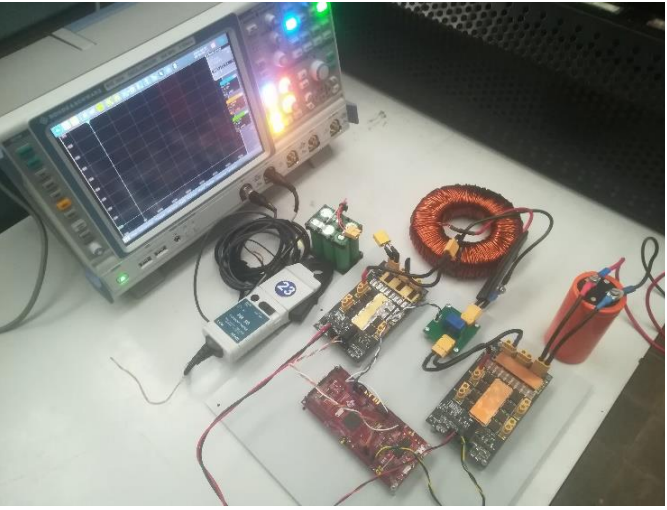

**Figure 3.** Experimental setup with custom ripple generator.

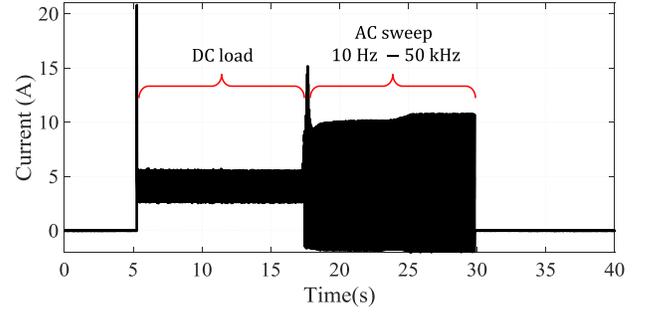

**Figure 4.** Measured load profile with initial DC load and a frequency sweep.

intercalation current is then proportional to the load current as follows

$$\hat{I}_{\text{int}} = \frac{1}{\sqrt{f_c^2 + f^2}} \hat{I}_{\text{load}}. \tag{18}$$

The ageing potential (15) consist of two load components: a DC component $I_{\text{DC}}$ equal to the average of the load current, and an AC component $\hat{I}_{\text{AC}}$ corresponding to the effective value of the AC component of the load current

$$j_{\text{ag}}^{AC} = \exp\left(-\frac{\alpha_{c,\text{ag}}F}{RT} R_{\text{ct}} \left(I_{\text{DC}} + \frac{1}{\sqrt{f_c^2 + f^2}} \hat{I}_{\text{AC}}\right)\right) \tag{19}$$

To separately extract the influence of the frequency on ageing, we introduce a so-called ageing potential as a proportion between the effects of pure DC load (where $\hat{I}_{AC} = 0$), and load with an AC component per

$$AP = \frac{j_{\text{ag}}^{AC}}{j_{\text{ag}}^{DC}} = \frac{k_{\text{ag}}^{AC}}{k_{\text{ag}}^{DC}} \exp\left(-\frac{\alpha_{c,\text{ag}}F}{RT} R_{\text{ct}} \frac{1}{\sqrt{f_c^2 + f^2}} \hat{I}_{\text{AC}}\right),$$

$$AP = A \exp\left(\frac{B}{\sqrt{C + f^2}}\right). \tag{20}$$

The ageing potential model (20) factorizes the contribution of an AC component of certain frequency to the reference DC load. The suggested model is further validated through measurement of specific cell properties and presented in following section.

## IV. EXPERIMENT

### A. Test setup

In the previous sections, we derived a simple ageing-potential model using combined cell modelling, which assumes high-pass characteristic of the cell impedance. We will further perform a battery measurement, fit the battery models presented in Section II, evaluate presumed attributes of the battery cell and estimate the cell ageing. Extracted data will be used to validate applicability of the suggested ageing-potential model (20).

We perform the measurement on a 6s module containing Sony/Murata US18650VTC5A 2600 mAh lithium-ion cells (Table I). The cells use a lithium nickel manganese cobalt oxide (NMC) cathode and a graphite anode. The data are measured at 80% SoC under an ambient temperature of 25°C.

The cell is loaded with a profile (see Fig. 4) which sweeps over frequencies ranging from 10 Hz up to 50 kHz. Rectangle pulses of the load profile are generated by custom electronics based on switched-inductor operation (see Fig. 3), where the output stage controls the magnitude and ripple of the inductor current. The test bench can reproduce load conditions of traction inverters. The load profile consists of initial 10 seconds of a DC current of 5 A to remove relaxation processes, which would affect further measurement (the DC character of the load is disturbed every 20 ms by switching the low side transistor for 30 μs to bootstrap the gate driver of the high side switches of the electronics). This initial phase is followed by increasing the amplitude up to 10 A while reducing the duty cycle down to 50 % (average current is kept at 5 A during the whole measurement). We measure the terminal voltage and the battery current at the battery terminals by 4-wire method. The current is sensed by a Hall-effect probe, where both current and voltage probes have a measurement bandwidth > 100 kHz.

TABLE I
BATTERY SPECIFICATION

| Load parameter | Value |
|---|---|
| Cell type | US18650VTC5A (Murata/Sony) |
| Nominal cell capacity | 2 600 mAh |
| Internal cell impedance | 10 mΩ |
| Cell max discharge current | 30 A |
| Battery configuration | 6s 1p |

TABLE II
BATTERY MODEL PARAMETERS

| Parameter | Value |
|---|---|
| Open circuit voltage, $V_{\text{OCV}}$ | 22 V |
| Exchange current, $i_{0,\text{int}}$ | 0.44 A |
| Charge transfer coefficient | 0.5 |
| Battery resistance, $R_0$ | 77.5 mΩ |
| Battery inductance, $L_0$ | 533 nH |
| SEI resistance, $R_{\text{SEI}}$ | 67 mΩ |
| SEI capacitance, $C_{\text{SEI}}$ | 23 mF |
| Electrode capacitance, $C_{\text{dl}}$ | 2.6 mF |
| Diffusion resistance, $R_{W1}$ | 0.6 mΩ |
| Diffusion capacitance, $C_{W1}$ | 3.5 mF |
| Diffusion resistance, $R_{W2}$ | 30 mΩ |
| Diffusion capacitance, $C_{W2}$ | 258 F |
| Temperature, $T$ | 298.15 K |
| Gas constant, $R$ | 8.314462 J K$^{-1}$ mol$^{-1}$ |
| Faraday constant, $F$ | 96 485.33 C mol$^{-1}$ |

*A. Battery modeling and parameter estimation*

The measurement data are used to identify parameters of the battery model of Section II, where the Warburg impedance is approximated with two R–C pairs. Rather than a charge transfer resistance only, the intercalation processes are represented by the full Butler–Volmer (Equation 3) to avoid limitations of the approximation. The fidelity of the model can be reviewed in Fig. 5, which compares measured and simulated voltage responses at various frequencies.

Identified parameters are displayed in Table II, and the model is further used to simulate over-potentials at loads of various frequencies. The identified battery model is experimentally loaded with a set of sinusoidal load profiles (Fig. 5), and the over-potential at the electrode–electrolyte interface is used to identify kinetics of the side reactions described in Section II.b. Loss of the lithium inventory following from Eqs. (9)–(12) is compared to a reference experiment with purely DC component. The ageing potential is simulated for frequencies ranging from 1 Hz up to 100 kHz and further used to validate our proposed ageing potential model.

## V. RESULTS

In addition to the ageing potential, the model also offers insights into the frequency influence on the battery loss and the source impedance of the cell. The AC impedance of the cell is lower than pure DC until the inductance starts limiting thanks to the shunting capability of the electrode capacitance (see Figure 6). Nevertheless, the impedance rapidly rises at higher frequencies as the cell inductance starts to dominate. Further, the cell exhibits increased losses related to an increased root mean square (RMS) current with superimposed AC component. Interestingly, the losses are progressively reduced at higher frequencies due to the decreased impedance of the cell. Similarly, the model explains the known drop of battery ageing potential at higher frequencies [16, 31-35]. The reduced ageing can be widely attributed to the shunting effect of the electrode capacitance, which is translated to decreased values of the over-potential. At higher frequencies, values of the over-potential are effectively filtered down to average values, which comparable to the DC load and so is the ageing at these frequencies.

We further use the data of the relative ageing potential to evaluate the feasibility of our simplified model (20). The parameters of the model are identified using least-square regression. The results of our simplified approximation are compared to the extracted values of ageing potential in Figure 7. The resultant ageing potential function can be expressed as

$$AP = 1.93 \exp\left(\frac{5.16 \cdot 10^6}{\sqrt{6.79 \cdot 10^5 + f^2}}\right), \quad (21)$$

where the R-squared coefficient of our further simplified ageing-potential approximation function (21) is $R^2 = 0.995$. Our simplified model fits well both low– and high–frequency regions and features a slightly sharper transition than the more detailed analytical model.

## VI. CONCLUSION

We introduced an ageing-potential model for the evaluation of ageing under rippled load. The model is derived from low-level chemistry models, from which we extracted the core of our ageing-potential model. The model was further conveniently combined with equivalent circuit modeling, which brought about the possibility to identify the model and calibrate it to a cell.

We further forked off a highly simplified regression model that catches the asymptotic behavior and allows very simple regression. The complexity of the model is on a similar level as the conventional Arrhenius model and offers practical usability. Despite all simplifications, the suggested model features high fidelity with R-squared coefficient > 0.99 with respect to data extracted from cell measurements and chemical modeling.

The suggested model may serve for benchmarking various battery cells in applications with inherently high load ripple

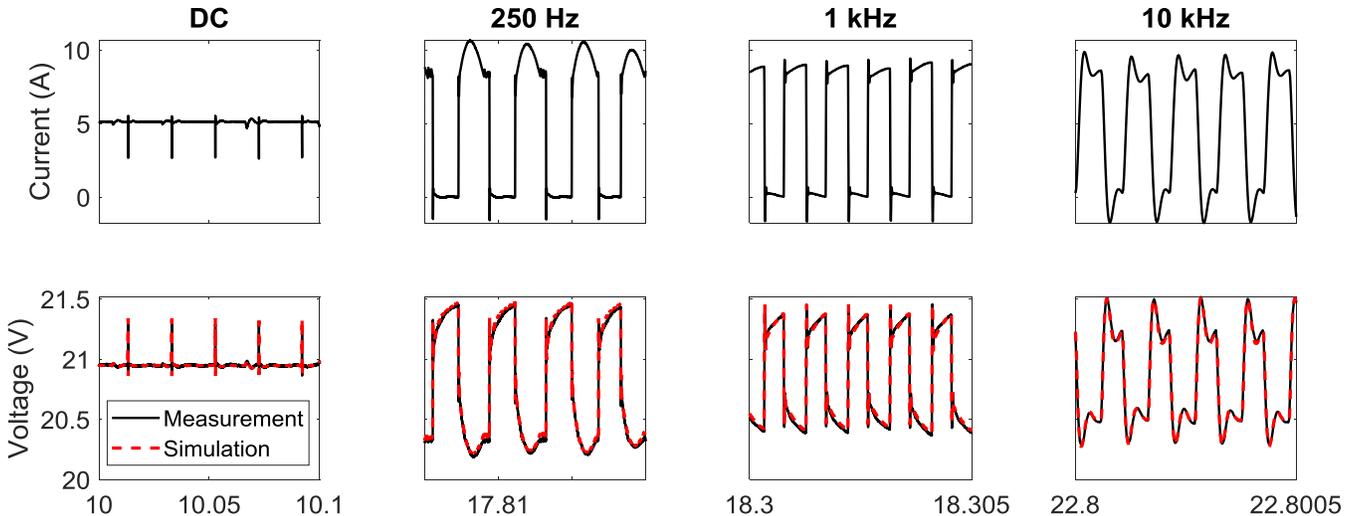

**Figure 5.** Synopsis of the regression model and measurement

(e.g., battery-based modular multilevel converters, electric vehicle drive trains with reduced dc–link capacitors), appropriate selection of switching frequency, and control strategies, as well as proper design of DC-link capacitors or input filter for battery interfaced inverters.

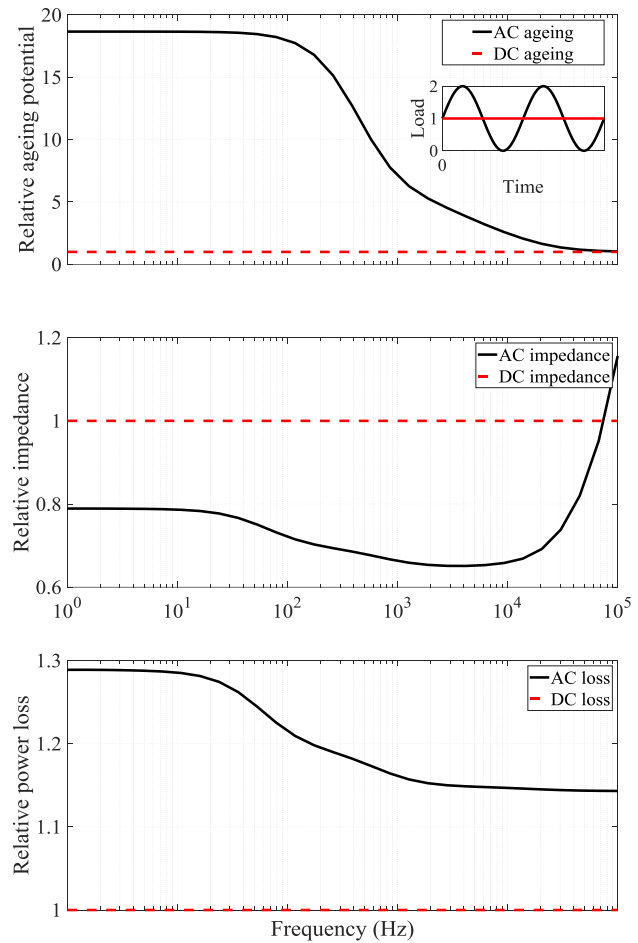

**Figure 6.** Results of the combined modeling approach. The model exploits ageing potential under rippled condition together with power loss and source impedance of examined cell.